\documentclass[nofootinbib,a4paper,12pt]{article}
\usepackage{jheppub}
\usepackage[T1]{fontenc}
\usepackage[utf8]{inputenc}
\usepackage{float}
\usepackage{slashed}
\usepackage{multicol,multirow}
\usepackage{amssymb}
\usepackage{amsmath}
\usepackage{subcaption}
\usepackage{array}
\usepackage{color}
\usepackage{hyperref}
\hypersetup{colorlinks=true}
\usepackage{longtable}
\usepackage{graphics,graphicx,epsfig,ulem,booktabs}
\numberwithin{equation}{section}

\definecolor{blue}{rgb}{0.2, 0.4, 1.0}
\definecolor{green}{rgb}{0.1,0.8,0.2}
\definecolor{orange}{rgb}{1.0,0.5,0.0}
\definecolor{cyan}{rgb}{0.0,0.75,0.8}

\newcolumntype{C}[1]{>{\centering\let\newline\\\arraybackslash\hspace{0pt}}m{#1}}

\newcommand{\GeV}{\, {\rm GeV}}

\title{\boldmath 
Two body non-leptonic $D^0$ decays 
from LCSR and implications for $\Delta a^{\rm dir}_{\rm CP}$
}

\preprint{SI-HEP-2023-34, P3H-23-105}

\author[a]{Alexander Lenz,}
\author[a]{Maria Laura Piscopo,}
\author[a]{Aleksey V. Rusov}

\affiliation[a]{Physik Department, Universit\"{a}t Siegen, 
Walter-Flex-Str. 3, 57068 Siegen, Germany}

\emailAdd{alexander.lenz@uni-siegen.de}
\emailAdd{maria.piscopo@uni-siegen.de}
\emailAdd{rusov@physik.uni-siegen.de}

\abstract{
Motivated by the recent measurements of CP violating effects in
singly Cabibbo suppressed $D^0$ decays, we revisit the theoretical predictions of these channels. Using up-to-date values for the decay constants and form factors, 
we find already within naive QCD factorisation
surprisingly good agreement between the central values of the branching ratios and the corresponding experimental data. We further extend the study of these modes by employing the method of light-cone sum rules~(LCSR) with light-meson light-cone distribution amplitudes. Using for the first time this framework to compute the leading contribution to the decay amplitude,
we can again describe well the experimental branching ratios for the modes 
$D^0 \to \pi^+ K^-$, $D^0 \to K^+ K^-$, 
$D^0 \to \pi^+ \pi^- $ and $D^0 \to K^+ \pi^-$.
The combination of our results with known predictions for the penguin contributions, obtained with LCSR, leads to an upper bound for the value of direct CP violation expected in the Standard Model of $|\Delta a_{\rm CP}^{\rm dir}| \leq { 2.4} \times 10^{-4} \,,$ which is approximately a factor six smaller than the current measurement.
}

\begin{document}

\maketitle

\section{Introduction}
Studies of charmed hadrons decays provide
complementary information with respect to the $b$-system on the structure of the Standard Model (SM), see e.g.~Ref.~\cite{Lenz:2020awd} for a review.
From a theoretical point of view, however, a robust  description of the charm sector is currently more challenging
due to the smaller value of the charm quark mass and, correspondingly, the larger 
size of the strong coupling at this scale.
In spite of this, it was recently found that
lifetimes of charmed mesons and baryons can be consistently computed within the heavy quark expansion (HQE), yielding results in agreement~\cite{King:2021xqp,Gratrex:2022xpm,Cheng:2023jpz} with the experimental data
\cite{HFLAV:2022esi} within uncertainties. This framework is currently quite advanced, with NLO-QCD corrections to the spectator quark contributions~\cite{Beneke:2002rj,Franco:2002fc,Lenz:2013aua},
as well as the coefficient at LO-QCD of the Darwin
operator~\cite{Lenz:2020oce,Mannel:2020fts,Moreno:2020rmk} known by now and, in the case of charmed mesons, also three-loop sum rule estimates for the matrix element of the arising four-quark dimension-six operators available~\cite{Kirk:2017juj,King:2021jsq} - in addition, first steps towards a determination of these non-perturbative inputs using Lattice QCD are being taken~\cite{Black:2023vju}.
On the other hand, the applicability of the HQE
to mixing observables is plagued by the presence of
an extreme GIM~\cite{Glashow:1970gm} suppression, making the theoretical interpretation of these quantities still not 
fully understood. Specifically, analyses based on simplified investigations of exclusive decays \cite{Falk:2001hx,Falk:2004wg}, i.e.\ taking only phase space effects into account but no dynamical QCD contributions, lead to a range of values for the mixing parameters that is consistent with the experimental data~\cite{HFLAV:2022esi}, whereas studies computed within the HQE yield extremely suppressed results, see e.g.~Ref.~\cite{Bobrowski:2010xg}.
Despite recent progress made in assessing the uncertainty
of the HQE prediction~\cite{Lenz:2020efu}, it is still unclear how to deal with such a strong GIM suppression from a theoretical point of view.
\\
Another peculiarity of the charm system is that CP violation effects are expected to be tiny, see e.g.~Ref.~\cite{Lenz:2020awd}.
Therefore, it came as a big surprise when in 2011 \cite{LHCb:2011osy} the LHCb Collaboration found evidence of sizable CP violating effects in singly Cabibbo suppressed decays of neutral $D$ mesons into two pions or two kaons.
The quantity considered is the difference of two time-integrated CP asymmetries, i.e. 
\begin{equation}
\Delta A_{\rm CP}  \equiv  
A_{\rm CP} (K^+ K^-) 
-
A_{\rm CP} (\pi^+ \pi^-) \, ,
\end{equation}
with the time dependent asymmetry defined as
\begin{eqnarray}
A_{\rm CP} (f; t) 
& = & 
\frac{
\Gamma (D^0 (t) \to f) - \Gamma (\overline{D}^0 (t) \to f)
}
{
\Gamma (D^0 (t) \to f) + \Gamma (\overline {D}^0 (t) \to f)
}
 \, ,
 \label{eq:Acp}
\end{eqnarray}
for an arbitrary final state $f$. The experimental discovery 
was finally made by the LHCb Collaboration in 2019~\cite{LHCb:2019hro}.
As the dominant contribution to Eq.~\eqref{eq:Acp} comes from direct CP violation, in the following we will consider only $\Delta a_{\rm CP}^{\rm dir}$. 
The current experimental average reads \cite{LHCb:2019hro}
\begin{equation}
\Delta a_{\rm CP}^{\rm dir}\big|_{\rm exp}   =
(-15.7 \pm 2.9) \times 10^{-4} \, ,
\label{eq:DeltaACP_exp}
\end{equation}
while a comprehensive summary of the evolution of both the
experimental and theoretical determinations for this observable can be found in~Ref.~\cite{Lenz:2013pwa} (from 2013)
and in Ref.~\cite{Chala:2019fdb} (from 2019). Recently, the LHCb Collaboration also presented a new measurement of the CP asymmetry in the $D^0 \to K^+ K^-$ channel~\cite{LHCb:2022lry}.
The latter, combined with the result for $\Delta a_{\rm CP}^{\rm dir}$ in Eq.~\eqref{eq:DeltaACP_exp}, leads to the following values for the size of CP asymmetry in the two individual modes, namely
\begin{align}
a_{\rm CP}^{\rm dir} (K^+ K^-)|_{\rm exp} & =  
(7.7 \pm 5.7) \times 10^{-4} \, ,
\label{eq:DeltaACP_KK_exp}
\\[2mm]
a_{\rm CP}^{\rm dir} (\pi^+ \pi^-)|_{\rm exp} & =  
(23.2 \pm 6.1) \times 10^{-4} \, ,
\label{eq:DeltaACP_pipi_exp}
\end{align}
where Eq.~\eqref{eq:DeltaACP_pipi_exp} provides the first evidence of CP violation in a specific $D$-meson decay.\\
Exclusive hadronic decays of charmed hadrons pose further challenges to robust theoretical predictions and a wide range of theory results can be found in the literature.
From naive estimates, see e.g.\ Ref.~\cite{Grossman:2006jg}, the value of $\Delta a_{\rm CP}^{\rm dir}$ is expected to be approximately one order of magnitude smaller than the one in Eq.~\eqref{eq:DeltaACP_exp}. 
This result was also confirmed in Ref.~\cite{Khodjamirian:2017zdu} using the framework of light-cone sum rule~(LCSR)~\cite{Balitsky:1989ry} with pion and kaon light-cone distribution amplitudes~(LCDAs), and similar conclusions were obtained in a recent analysis of final state interactions employing dispersion relations~\cite{Pich:2023kim}.
Consequently, the large experimental value in Eq.~\eqref{eq:DeltaACP_exp} has also triggered several investigations of physics beyond the SM~(BSM)~\cite{Chala:2019fdb,Dery:2019ysp,Calibbi:2019bay,Bause:2020obd}.\\
At the same time, in the literature there are also theory estimates that point towards a SM origin of the experimental result for $\Delta A_{\rm CP}$.
These include analyses based on U-spin relations, see e.g.\ Ref.~\cite{Grossman:2019xcj}, studies of rescattering contributions with potential large enhancements due 
to nearby resonances like the $f_0(1710)$ or $f_0(1790)$~\cite{Schacht:2021jaz}, as well as investigations of final state interactions~\cite{Bediaga:2022sxw}. On the other hand, the effect of nearby resonances is in principle also included in the approach of Ref.~\cite{Pich:2023kim}, where no sign of large enhancement was found and, furthermore, Ref.~\cite{Pich:2023kim} points out some inconsistencies e.g.\ in Ref.~\cite{Bediaga:2022sxw}. In addition, also approaches based on topological diagram analyses have been employed~\cite{Li:2019hho,Cheng:2019ggx,Wang:2020gmn}. In particular, in Ref.~\cite{Cheng:2019ggx} the experimental data were combined with certain theory assumptions, including ad-hoc guesses on the relative size of 
the QCD-penguin exchange graphs, in order to estimate the different topological contributions, clearly not constituting a first principle determination. Finally, it is also worth commenting that the new experimental results 
for the individual CP asymmetries 
shown in Eqs.~\eqref{eq:DeltaACP_KK_exp}, \eqref{eq:DeltaACP_pipi_exp}, would imply a huge U-spin symmetry breaking~\cite{Schacht:2022kuj, Bause:2022jes}. Currently then, a clear theory interpretation of the data is still missing. \\
In Ref.~\cite{Khodjamirian:2017zdu}, Khodjamirian and Petrov used LCSR to estimate the size of penguin contributions in $D^0 \to \pi^+ \pi^-$ and $D^0 \to K^+ K^-$, whereas the value of the leading decay amplitude, needed to predict $\Delta a_{\rm CP}^{\rm dir}$, see Section~\ref{sec:theory-framework}, was extracted from  experimental data on the corresponding branching fractions. The latter are known quite precisely, and the PDG quotes~\cite{ParticleDataGroup:2022pth}
\begin{align}
{\cal B} (D^0 \to K^+ K^-)\big|_{\rm exp}    & =  (4.08 \pm 0.06) 
\times 10^{-3} \, ,
\label{eq:Br_exp-KK}
\\[2mm]
{\cal B} (D^0 \to \pi^+ \pi^-)\big|_{\rm exp}  & =  (1.454 \pm 0.024) 
\times 10^{-3} \, .
\label{eq:Br_exp-pipi}
\end{align}
The result obtained for the penguin amplitude was approximately one order of magnitude smaller than what would be required to explain the value of $\Delta a_{\rm CP}^{\rm dir}$ in Eq.~\eqref{eq:DeltaACP_exp}. To gain further insight into the question whether this difference is an artefact of the method used or a signal of tension between the SM prediction and the data, 
in this paper we present a new determination of the leading contribution to the decay amplitude of two-body singly Cabibbo suppressed $D^0$ decays using exactly the same framework as done in Ref.~\cite{Khodjamirian:2017zdu}, and compare with the corresponding experimental values in Eqs.~\eqref{eq:Br_exp-KK}, \eqref{eq:Br_exp-pipi}.
In addition, we also extend our analysis to the Cabibbo favoured and doubly Cabibbo suppressed decays $D^0 \to \pi^+ K^- $ and $D^0 \to K^+ \pi^-$, for which the experimental results of the branching fractions read~\cite{ParticleDataGroup:2022pth} 
\begin{align}
{\cal B} (D^0 \to \pi^+ K^-)\big|_{\rm exp} & = (3.947 \pm 0.030) 
\times 10^{-2} \, ,
\label{eq:Br_exp-Kpi}
\\[2mm]
{\cal B} (D^0 \to K^+ \pi^-)\big|_{\rm exp} & = (1.50 \pm 0.07) 
\times 10^{-4} \, .
\label{eq:Br_exp-piK}
\end{align}
Obtaining a good agreement would considerably strengthen our confidence in the applicability of LCSR for two-body non-leptonic $D^0$ meson decays, and represent a first crucial step towards a more robust understanding of the strong dynamics in these channels. 
\\
Our paper is organised as follows: 
in Section~\ref{sec:theory-framework} we introduce the theoretical framework needed to describe branching ratios and CP asymmetries, starting from the effective Hamiltonian. In Section~\ref{sec:nQCDf} we revisit the naive QCD factorisation estimates for the branching fractions, using  updated theoretical and experimental inputs. The main result of our paper, the calculation of the tree-level amplitude from LCSR, is described in Section~\ref{sec:LCSR}.
Section~\ref{sec:Result} is devoted to the description of the numerical analysis and the discussion of our results, including predictions for the branching fractions and their implications on the bound on $|\Delta a_{\rm CP}^{\rm dir}|$. Finally, in Section~\ref{sec:Conclusion} we conclude and discuss future potential improvements to our calculation.
\section{The general framework}
\label{sec:theory-framework}
\subsection{Effective Hamiltonian and decay amplitudes}
The singly Cabibbo suppressed decays~\footnote{Note that the description can be easily generalised to the Cabibbo favoured and doubly Cabibbo suppressed decays $D^0 \to \pi^+ K^-$ and $D^0 \to K^+ \pi^-$. We omit this for brevity.} $D^0 \to K^+ K^-$, $D^0 \to \pi^+ \pi^-$, can be described by introducing the $\Delta C = 1$ effective Hamiltonian governing the flavour-changing charm-quark transitions $c \to q \bar q u$, with $q = d, s$, namely~\cite{Buchalla:1995vs} 
\begin{align}
  {\cal H}_{\rm eff}  =  
  \frac{G_F}{\sqrt{2}} 
  \sum_{q = d,s} \lambda_{q} \Bigl(  C_1 O_1^{q} +
  C_2 O_2^{q} \Bigr)  + {\rm h.c.}\,,
   \label{eq:Heff}
\end{align}
where $\lambda_{q} \equiv V^*_{cq} V_{uq}$ 
are the elements of the Cabibbo–Kobayashi–Maskawa~(CKM) matrix, and the current-current operators $O_{1,2}^{q}$ read respectively
\begin{equation}
    O_1^{q} = (\bar q^i \Gamma_\mu c^i) (\bar u^j \Gamma^\mu q^j)\,, \qquad 
    O_2^{q} = (\bar q^i \Gamma_\mu c^j) (\bar u^j \Gamma^\mu q^i)\,.
    \label{eq:O1-O2}
\end{equation}
Here $\Gamma_\mu \equiv \gamma_\mu(1-\gamma_5)$, $i,j,$ are colour indices, and the corresponding Wilson coefficients $C_i(\mu_1)$ are evaluated at the renormalisation scale $\mu_1 \sim m_c$. Note that in Eq.~\eqref{eq:Heff} the contribution of the penguin and chromomagnetic operators has been neglected due to the smallness of their Wilson coefficients compared to the accuracy of our study.\\
Similarly to Ref.~\cite{Khodjamirian:2017zdu}, we also introduce a compact notation for the combination of the effective operators $O^{q}_{1,2}$ and of their Wilson coefficients in Eq.~\eqref{eq:Heff}, i.e.
\begin{equation}
    {\cal O}^{q} \equiv - \frac{G_F}{\sqrt 2}\sum_{i=1,2}C_i O_i^{q}\,, \quad \mbox{with } q = d,s\,,
\label{eq:Oq-def}
\end{equation}
so that each decay amplitude can be expressed in terms of the corresponding CKM structure respectively as
\begin{align}
    {\cal A}(D^0 \to \pi^+ \pi^-) &= \lambda_d \langle \pi^+ \pi^-| {\cal O}^d| D^0 \rangle + \lambda_s \langle \pi^+ \pi^-| {\cal O}^s| D^0 \rangle \,,
    \label{eq:Am-pipi}
    \\[2mm]
   {\cal A}(D^0 \to K^+ K^-) &= \lambda_s \langle K^+ K^-| {\cal O}^s| D^0 \rangle + \lambda_d \langle K^+ K^-| {\cal O}^d| D^0 \rangle \,.
    \label{eq:Am-KK}
\end{align}
\begin{figure}[t]
    \begin{minipage}{0.48\textwidth}
    \centering
    \includegraphics[scale=0.43]{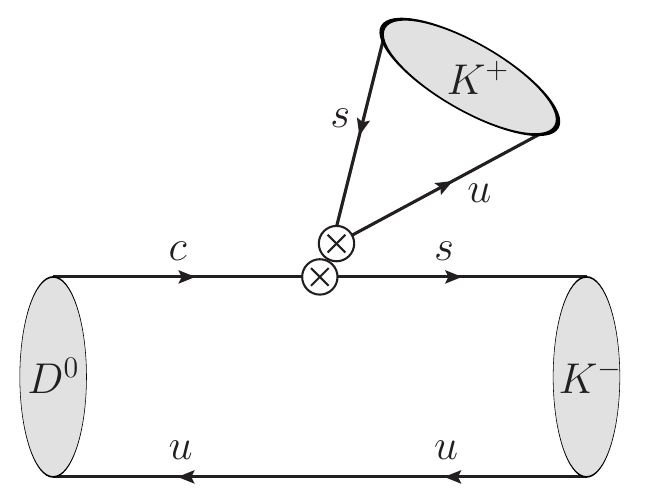}\\
    (a)
    \end{minipage}
    \begin{minipage}{0.48\textwidth}
    \centering
    \includegraphics[scale=0.43]{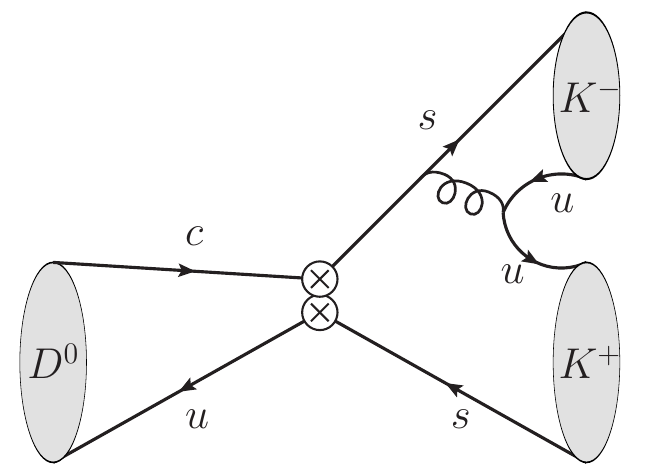}\\
    (b)
    \end{minipage}
   \begin{minipage}{0.48\textwidth}
    \centering
    \includegraphics[scale=0.43]{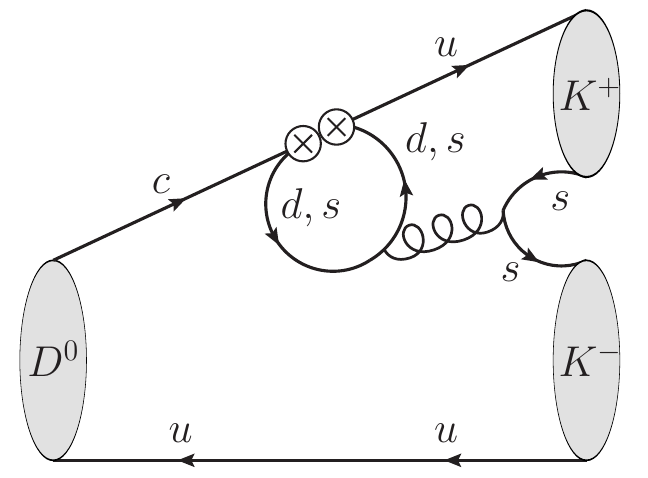}\\
    (c)
    \end{minipage}
    \begin{minipage}{0.48\textwidth}
    \centering
    \includegraphics[scale=0.43]{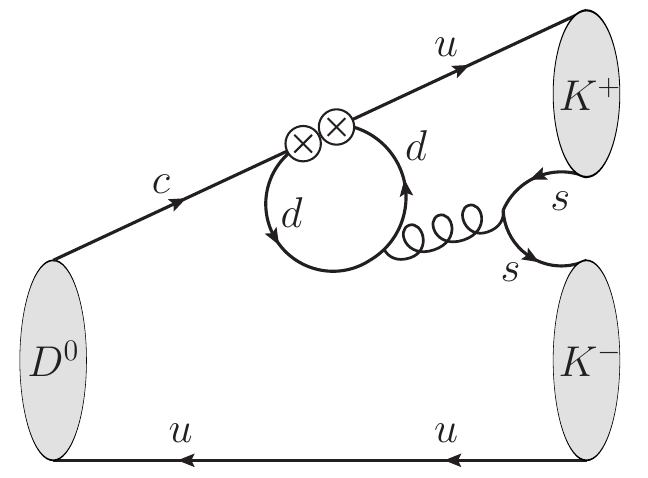}\\
    (d)
    \end{minipage}
    \caption{Examples of tree-level (a), exchange (b) and penguin (c) topologies contributing to ${\cal A}_{KK}$. Example of penguin topology contributing to ${\cal P}_{KK}$ (d).
    The corresponding diagrams for ${\cal A}_{\pi \pi}$ and ${\cal P}_{\pi \pi}$ can be obtained replacing $K \to \pi$, $s \leftrightarrow d$.}
    \label{fig:AKK-diagrams}
\end{figure}
Using the unitarity of the CKM matrix  $\lambda_d + \lambda_s + \lambda_b = 0$, Eqs.~\eqref{eq:Am-pipi}, \eqref{eq:Am-KK}, are then recast in the following form
\begin{align}
    {\cal A}(D^0 \to \pi^+ \pi^-) & =  \lambda_d \, {\cal A}_{\pi \pi} \left[1 - \frac{\lambda_b}{\lambda_d} \frac{{\cal P}_{\pi\pi}}{{\cal A }_{\pi \pi}} \right] \,,
    \label{eq:A-Dpipi}
    \\[2mm]
   {\cal A}(D^0 \to K^+ K^-) &= \lambda_s \, {\cal A}_{K K} \left[1 - \frac{\lambda_b}{\lambda_s} \frac{{\cal P}_{KK}}{{\cal A }_{KK}} \right]\,,
    \label{eq:A-DKK}
\end{align}
where we have defined, respectively
\begin{align}
{\cal A}_{\pi\pi} & = \langle \pi^+ \pi^-| {\cal O}^d| D^0 \rangle - \langle \pi^+ \pi^-| {\cal O}^s| D^0 \rangle\,,
 \label{eq:Apipi}
\\[2mm]
{\cal A}_{KK} & = \langle K^+ K^-| {\cal O}^s| D^0 \rangle - \langle K^+ K^-| {\cal O}^d| D^0 \rangle\,,
 \label{eq:Akk}
\end{align}
and 
\begin{equation}
    {\cal P}_{\pi\pi} = \langle \pi^+ \pi^-| {\cal O}^s| D^0 \rangle\,, \qquad   {\cal P}_{KK} = \langle K^+ K^-| {\cal O}^d| D^0 \rangle\,.
    \label{eq:P}
\end{equation}
The leading CKM amplitudes ${\cal A}_{\pi\pi}$, ${\cal A}_{KK}$, in Eqs.~\eqref{eq:Apipi}, \eqref{eq:Akk}, receive contributions from color-allowed tree-level, exchange and penguin topologies, on the other hand, only the penguin topology can contribute to ${\cal P}_{\pi\pi}$, ${\cal P}_{KK}$, in Eq.~\eqref{eq:P}, cf.\  Fig.~\ref{fig:AKK-diagrams}. 
\subsection{Branching fractions and CP asymmetries}
The branching ratio for the decay $D^0 \to K^+ K^-$ can be then expressed as
\begin{equation}
   {\cal B}(D^0 \to K^+ K^-) = {\cal N}_{KK} |\lambda_s|^2 |{\cal A}_{KK}|^2 \left|1 - \frac{\lambda_b}{\lambda_s} \frac{{\cal P}_{KK}}{{\cal A}_{KK}} \right|^2 \,,
   \label{eq:BR-KK}
 \end{equation}
where the prefactor ${\cal N}_{KK}$ is
 \begin{equation}
   {\cal N}_{KK} \equiv  \frac{ \sqrt{\lambda(m_D^2,m_K^2, m_K^2)} }{16 \pi m_D^3} \, \tau(D^0)\,,
   \label{eq:N_KK}
\end{equation}
with $\lambda(a,b,c) \equiv (a - b - c)^2 - 4 bc $ being the K\"allen function and $\tau(D^0)$ the total lifetime of the $D^0$ meson. Similarly, the direct CP asymmetry, defined as
\begin{equation}
a_{\rm CP}^{\rm dir}(f) \equiv \frac{
\Gamma (D^0 \to f) - \Gamma (\overline{D}^0 \to \bar f)
}
{
\Gamma (D^0 \to f) + \Gamma (\overline {D}^0  \to \bar f)
}\,,
\end{equation}
reads
\begin{equation}
    a_{\rm CP}^{\rm dir}(K^+ K^-) = - \frac{\displaystyle 2 \left|\frac{\lambda_b}{\lambda_s}  \right| \sin{\gamma} \left| \frac{{\cal P}_{KK}}{{\cal A}_{KK}} \right| \sin{\phi}_{KK} }{\displaystyle 1-2 \left|\frac{\lambda_b}{\lambda_s} \right| \cos{\gamma} \left| \frac{{\cal P}_{KK}}{{\cal A}_{KK}} \right| \cos{\phi}_{KK}  + \left|\frac{\lambda_b}{\lambda_s} \right|^2  \left| \frac{{\cal P}_{KK}}{{\cal A}_{KK}} \right|^2  }\,,
    \label{eq:acp_dir}
\end{equation}
where we have defined the strong phase difference $\phi_{KK} \equiv \arg \left({\cal P}_{KK}/{\cal A}_{KK} \right)$, and introduced the angle $\gamma \equiv -\arg(\lambda_b/\lambda_s)$. 
Note that the corresponding expressions for the mode $D^0 \to \pi^+ \pi^-$ can be obtained by replacing $K K \to \pi \pi$, $\lambda_s \to \lambda_d$ and $\sin \gamma \to - \sin \gamma$ in Eqs.~\eqref{eq:BR-KK}, \eqref{eq:N_KK}, and \eqref{eq:acp_dir}. \\
Taking into account that $\lambda_b/\lambda_{d,s} \ll 1$, it thus follows that while the amplitudes ${\cal A}_{\pi\pi}$, ${\cal A}_{KK}$, give the dominant contribution to the branching fractions, i.e. 
\begin{align}
 {\cal B}(D^0 \to \pi^+ \pi^-) &\simeq {\cal N}_{\pi\pi}|\lambda_d|^2 |{\cal A}_{\pi \pi}|^2 \,,
\label{eq:BR-D0-to-pi-pi}
\\[2mm]
{\cal B}(D^0 \to K^+ K^-)&\simeq {\cal N}_{KK} |\lambda_s|^2 |{\cal A}_{KK}|^2\,,
\label{eq:BR-D0-to-K-K}    
\end{align}
the direct CP asymmetries are driven by the ratio of the penguin over the CKM leading amplitudes, that is
\begin{align}
    a_{\rm CP}^{\rm dir}(\pi^+ \pi^-) &\simeq 2 \left|\frac{\lambda_b}{\lambda_d}  \right| \sin{\gamma} \left| \frac{{\cal P}_{\pi\pi}}{{\cal A}_{\pi\pi}} \right| \sin \phi_{\pi\pi}  \,, 
    \label{ew:a_pipi_appr}
\\[2mm]
    a_{\rm CP}^{\rm dir}(K^+ K^-) &\simeq  - 2 \left|\frac{\lambda_b}{\lambda_s}  \right| \sin{\gamma} \left| \frac{{\cal P}_{KK}}{{\cal A}_{KK}} \right| \sin \phi_{KK}\,.
    \label{ew:a_KK_appr}
\end{align}
From the above results, together with $|\lambda_d| \simeq |\lambda_s|$, we arrive at the following expression for the difference of direct CP asymmetries $\Delta a_{\rm CP}^{\rm dir}$, that is
\begin{equation}
\Delta a_{\rm CP}^{\rm dir} \simeq - 2 \left|\frac{\lambda_b}{\lambda_s} \right| \sin{\gamma} 
\left(\left| \frac{{\cal P}_{KK}}{{\cal A}_{KK}} \right| \sin \phi_{KK} 
+ \left| \frac{{\cal P}_{\pi\pi}}{{\cal A}_{\pi\pi}} \right| \sin \phi_{\pi\pi}\right) \,.
\label{eq:Delta-ACP}
\end{equation}
The penguin amplitudes ${\cal P}_{\pi\pi}$, ${\cal P}_{KK}$, were estimated in Ref.~\cite{Khodjamirian:2017zdu} using the framework of LCSR with respectively pion and kaon LCDAs, and  following previous studies of the $B \to \pi \pi$ decay~\cite{Khodjamirian:2000mi, Khodjamirian:2003eq}. 
The values of $|{\cal A}_{\pi \pi}|$ and $|{\cal A}_{KK}|$, required to determine $a_{\rm CP}^{\rm dir}$, were instead extracted from the precise experimental data available on the branching ratios, taking into account the relations in Eqs.~(\ref{eq:BR-D0-to-pi-pi}), (\ref{eq:BR-D0-to-K-K}). At the same time, assuming naive power counting, we can express ${\cal A}_{\pi \pi}$ and ${\cal A}_{KK}$ in Eqs.~\eqref{eq:Apipi}, \eqref{eq:Akk}, as
\begin{align}
    {\cal A}_{\pi \pi} &=  \langle \pi^+ \pi^-| {\cal O}^d| D^0 \rangle \Bigl|_{\rm tree} + \, {\cal O}(\alpha_s) + {\cal O}(1/m_c)\,,
    \\[2mm]
    {\cal A}_{K K} &=  \langle K^+ K^-| {\cal O}^s| D^0 \rangle \Bigl|_{\rm tree} + \, {\cal O}(\alpha_s) + {\cal O}(1/m_c)\,,
\end{align}
retaining the dominant contribution due to the tree-level amplitude and neglecting sub-leading diagrams due to both hard and soft QCD corrections. In this approximation it thus follows that 
\begin{align}
{\cal A_{\pi \pi}} &\simeq -\frac{G_F}{\sqrt 2}  \left(C_1 + \frac{C_2}{3} \right)    
\langle \pi^+ \pi^- | O_1^{d} | D^0 \rangle \Bigl|_{\rm tree}\,,
\label{eq:Apipi-appr}
\\[2mm]
{\cal A}_{K K} &\simeq -\frac{G_F}{\sqrt 2}   \left(C_1 + \frac{C_2}{3} \right)    
\langle K^+ K^- | O_1^s | D^0 \rangle \Bigl|_{\rm tree}\,.
\label{eq:AKK-appr}
\end{align}
The tree-level matrix elements in Eqs.~\eqref{eq:Apipi-appr}, \eqref{eq:AKK-appr}, can be then estimated using naive QCD factorisation (nQCDf), see Section~\ref{sec:nQCDf}, or determined from a LCSR computation, see Section~\ref{sec:LCSR}. In the latter case, in fact, the combination with the results for ${\cal P}_{\pi \pi}$, ${\cal P}_{KK}$, obtained in Ref.~\cite{Khodjamirian:2017zdu}, allows us to consistently estimate the ratios in Eqs.~\eqref{ew:a_pipi_appr}, \eqref{ew:a_KK_appr}, and to ultimately obtain a constraint on $|\Delta a_{\rm CP}^{\rm dir}|$ entirely within the same theoretical framework. Note that for brevity, the suffix ``tree'' will be omitted in the following, although this should always be understood.\\
The computation of the tree-level hadronic matrix elements for the singly Cabibbo suppressed $D^0$ decays can be easily extended to the Cabibbo favourite and doubly Cabibbo suppressed decays $D^0 \to \pi^+ K^-$ and $D^0 \to K^+ \pi^-$,
which are also triggered by color-allowed tree-level topologies. Therefore, we include these two additional channels in our analysis. As for the color-suppressed tree-level decays like $D^0 \to \pi^0 \pi^0$, the prefactor $(C_1 + C_2/3)$ is replaced by $(C_2 + C_1/3)$, where the latter combination shows an almost perfect cancellation. 
This makes the analysis of these decays extremely sensitive to the accuracy of the computation, and hence we leave the inclusion of these channels to future studies that will also take higher order perturbative contributions into account.
\section{The decay amplitudes in naive QCDf}
\label{sec:nQCDf}
A first estimate of the tree-level hadronic matrix elements in Eqs.~\eqref{eq:Apipi-appr}, \eqref{eq:AKK-appr}, can be obtained within nQCDf. Considering, for instance, the decay $D^0 \to K^+ K^-$, the nQCDf approximation leads to
\begin{equation}
\langle K^+ K^- |O_{1}^{s} |D^0 \rangle \Bigl|_{\rm nQCDf} =
i f_{K} (m_D^2 - m_{K}^2) f_0^{D K} (m_{K}^2)\,,
\label{eq:ME-NQCDF}
\end{equation}
where $f_{K}$ is the kaon decay constant and $f_0^{D K} (m_{K}^2)$ the scalar form factor evaluated at $q^2 = m_{K}^2$. Hence, the amplitude ${\cal A}_{KK}$ becomes
\begin{equation}
{\cal A}_{KK}\Bigl|_{\rm nQCDf} = -i \frac{G_F}{\sqrt 2} \left(C_1 + \frac{C_2}{3} \right) f_K (m_D^2 - m_K^2) f_0^{D K} (m_{K}^2)\,.
\label{eq:AKK-NQCDF}
\end{equation}
Similar expressions for the remaining modes can be easily obtained by properly replacing $f_K \to f_\pi$, $f_0^{D K} (m_{K}^2) \to f_0^{D \pi} (m_{\pi}^2) \simeq f_0^{D \pi} (0)$, etc.\\
Using for all the relevant parameters like Wilson coefficients, masses, decay constants, etc.,
the values presented in Section~\ref{sec:Result}, see also Table~\ref{tab:input}, 
and in addition, Lattice QCD determinations for the form factors~\cite{Lubicz:2017syv}~\footnote{We use the most recent and so far only published Lattice QCD results using $N_f = 2 + 1 + 1$ ensembles~\cite{Lubicz:2017syv},
There are, in fact, also older Lattice QCD determinations based on using $N_f = 2 + 1$ ensembles~\cite{Na:2010uf, Na:2011mc}, which indicate slightly smaller SU(3)$_f$ breaking effects in the
form factors compared to Ref.~\cite{Lubicz:2017syv}. 
For more details, see the FLAG review~\cite{FLAG}.}, namely
\begin{align}
f_0^{D K} (0) & = 0.765 \pm 0.031\,,  \quad 
& f_0^{D K} (m_K^2) & = 0.789 \pm 0.028\,,  
\label{eq:FF_DK}
\\[2mm]
f_0^{D \pi} (0) & = 0.612 \pm 0.035\,,  \quad 
& f_0^{D \pi} (m_K^2) & = 0.639 \pm 0.032\,,
\label{eq:FF_Dpi}
\end{align}
we arrive at the following estimates for the branching fractions
\begin{eqnarray}
{\cal B} (D^0 \to K^+ K^-)\Big|_{\rm nQCDf} & = & \left(3.40^{+0.40}_{-0.35}\right) \times 10^{-3}\,,
\label{eq:BR-D0-to-K-K-nqcdf} \\[2mm]
{\cal B} (D^0 \to \, \pi^+ \, \pi^-)\Big|_{\rm nQCDf} & = & \left(1.90^{+0.28}_{-0.26}\right) \times 10^{-3}\,,
\label{eq:BR-D0-to-pi-pi-nqcdf} \\[2mm]
{\cal B} (D^0 \to \pi^+ K^- )\Big|_{\rm nQCDf} & = & \left(4.55^{+0.56}_{-0.50}\right) \times 10^{-2}\,,
\label{eq:BR-D0-to-K-pi-nqcdf} \\[2mm]
{\cal B} (D^0 \to K^+ \pi^-)\Big|_{\rm nQCDf} & = & \left(1.48^{+0.20}_{-0.19}\right) \times 10^{-4}\,,
\label{eq:BR-D0-to-pi-K-nqcdf}
\end{eqnarray}
where the quoted uncertainties include the variation of the input parameters, 
as well as of the renormalisation scale in the interval $1 \GeV \leq \mu_1 \leq 2 \GeV$, all combined in quadrature. We stress however, that these should not be understood as the final theory errors, since uncertainties due to contributions that cannot be captured by the nQCDf approximation are not included and could potentially be sizable.
\\
Surprisingly, already the nQCDf estimates are found to be in excellent agreement
with the corresponding data in Eqs.~(\ref{eq:Br_exp-KK}) - (\ref{eq:Br_exp-piK}). 
In particular, using updated input values for the form factors, we are able to reproduce the large experimental result for the SU(3)$_f$ breaking in the $D^0 \to K^+ K^-$ and $D^0 \to \pi^+ \pi^-$ modes. \\
To get a first idea of the possible size of sub-leading contributions to the decay amplitude, we consider the deviation of the experimental branching ratios from the nQCDf results in Eqs.~\eqref{eq:BR-D0-to-K-K-nqcdf} - \eqref{eq:BR-D0-to-pi-K-nqcdf}, as well as the corresponding deviation at the amplitude level in order to subtract the effect of simple phase space and CKM factors. Specifically, we define
\begin{align}
\delta {\cal B}^{\rm nQCDf} \equiv \frac{
({\cal B}^{\rm exp} - {\cal B}^{\rm nQCDf})}{{\cal B}^{\rm exp}} \,, \quad 
\delta {\cal A}^{\rm nQCDf} \equiv \frac{({\cal A}^{\rm exp} - {\cal A}^{\rm nQCDf})}{{\cal A}^{\rm exp}}\,,
\end{align}
obtaining the values summarised in the table below.
\begin{displaymath}
\renewcommand{\arraystretch}{1.3}
\begin{array}{|c||c|c|c|c|c|}
\hline
& D^0 \to K^+ K^-   & D^0 \to \pi^+ \pi^- 
& D^0 \to \pi^+ K^- & D^0 \to K^+ \pi^- \\
\hline
\hline
\delta {\cal B}^{\rm nQCDf} & 
0.17 & - 0.31 & -0.15 & 0.02
\\
\hline
\delta {\cal A}^{\rm nQCDf} & 
0.09 & -0.14 & -0.07 & 0.01
\\
\hline\end{array}
\end{displaymath}
These results demonstrate that the decay amplitudes for the channels considered are indeed dominated by the contribution of the tree-level topology, and that therefore there is no indication for a large enhancement of sub-leading diagrams.
It is important to stress that compared to previous analyses, like e.g.~Ref.~\cite{Grossman:2006jg}, several inputs have changed considerably. For instance, the $D \to \pi$ and $D \to K$ form factors
show now a significant SU(3)$_f$ breaking effect, cf.\ Eqs.~\eqref{eq:FF_DK}, \eqref{eq:FF_Dpi}, whereas the values from 2005 used in Ref.~\cite{Grossman:2006jg}
were almost SU(3)$_f$ symmetric. In addition, the experimental data for the branching ratios have also
changed, e.g.\ the value of $\Gamma (D^0 \to K^0 \bar K^0)$, which is expected to vanish in nQCDf, went down by more than a factor of five, while it seemed to be sizable in 2006, thus potentially contradicting the nQCDf estimate.
\section{The decay amplitudes from LCSR}
\label{sec:LCSR}
In this section, we outline the computation of the tree-level matrix elements in Eqs.~\eqref{eq:Apipi-appr}, \eqref{eq:AKK-appr}, using the framework of LCSR. For definiteness, we consider the decay $D^0 \to K^+ K^-$,
as it is more general. The calculation of the remaining channels is in fact analogous and the corresponding results can be obtained from those presented here implementing the proper replacements i.e.\ $K \to \pi$, $m_s \to m_d \to 0$, $m_K \to m_\pi \to 0$, etc.
Note that the analysis largely follows the studies of the $B \to \pi \pi$ and $B \to \pi K$ decays performed in Refs.~\cite{Khodjamirian:2000mi, Khodjamirian:2003xk}.\\
We start by introducing the three-point correlation function
\begin{equation}
F_{\mu} (p, q) = i^2 \int d^4 x \, e^{- i p \cdot x} \int d^4 y
\, e^{i q \cdot y} \,
\langle K^- (p - q)| {\rm T} \! \left\{j^{D}_5 (x), O_1^{s} (0), j^{K}_\mu (y) \right\} |0 \rangle\,,
\label{eq:Correlator-O1}
\end{equation}  
where $j^{K}_\mu =  \bar s \gamma_\mu \gamma_5 u$
and $j^{D^0}_5 = i  m_c \, \bar c \gamma_5 u $
are interpolating currents of the $K^+$ and $D^0$ mesons, respectively, with the corresponding off-shell momenta $p^\mu$ and~$q^\mu$. 
\begin{figure}[t]
    \centering
    \includegraphics[scale=0.6]{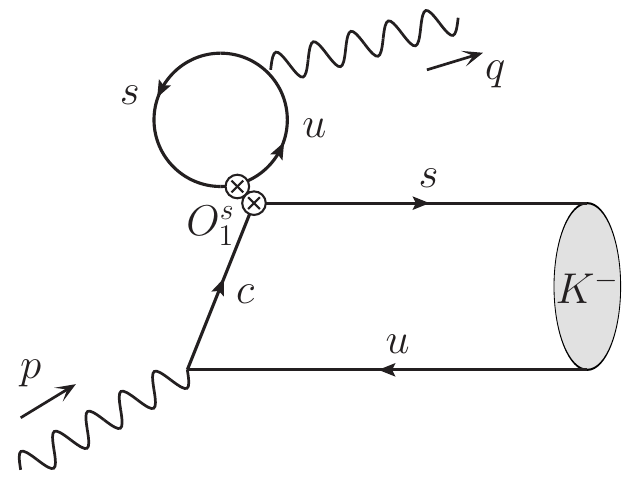}
    \caption{Diagram relevant for the computation of the LC-OPE for the the correlation function in Eq.~\eqref{eq:Correlator-O1}.}
    \label{fig:diagrams-OPE}
\end{figure}
The correlation function $F_\mu (p, q)$ admits the Lorentz decomposition
\begin{equation}
F_\mu(p,q) =  
F_q (p^2,q^2) \, q_\mu +
F_p (p^2,q^2) \, p_\mu \,,
\label{eq:Fq-Fp}
\end{equation}
in terms of the scalar functions $F_q (p^2,q^2)$ and $F_p (p^2,q^2)$,
depending on the two independent Lorentz invariant variables $p^2$ and $q^2$. We stress that
only the coefficient of $q_\mu$ contributes to the dispersion relations and is thus relevant for our analysis.\\
For sufficiently large space-like values of the momenta squared $P^2 \equiv - p^2 \gg \Lambda^2$, $Q^2 \equiv - q^2 \gg \Lambda^2$, with $\Lambda$ denoting a hadronic scale of the order of few hundreds MeV, one can show that the dominant contribution to the integrals in Eq.~\eqref{eq:Correlator-O1} comes from the light-cone region $x^2 \sim 0$ and $y^2 \sim 0$, see e.g.\ Ref.~\cite{Khodjamirian:2000mi}.
Therefore, with the kinematics fixed above, the so-called light-cone operator-product expansion (LC-OPE) is applicable, allowing us to represent the correlation function in the form of convolution of hard scattering kernels with the corresponding kaon LCDAs of growing twist. \\
The general expression for the non-local two-particle  kaon-to-vacuum matrix element expanded near the light-cone and up to twist-4, can be found e.g.\ in the Appendix of Ref.~\cite{Duplancic:2008ix}, and reads
\begin{eqnarray}
 \langle K^- (k)| \bar s^i_{\alpha} (x_1) u_{\beta}^j (x_2) |0 \rangle & = & \frac{i \delta^{ij}}{12} f_K \int\limits_0^1 e^{i u (k \cdot x_1) + i \bar u (k \cdot x_2)} \Bigl( [\slashed p \gamma_5]_{\beta \alpha} \phi_{2K} (u) 
\\
& & \hspace*{-10mm} - \, [\gamma_5]_{\beta \alpha} \mu_K \phi_{3 K}^p (u) + \frac{1}{6} [\sigma_{\mu \nu} \gamma_5]_{\beta \alpha} k^{\mu} (x_1 - x_2)^\nu \mu_K \phi_{3 K}^\sigma (u) 
\nonumber \\
& & \hspace*{-10mm} + \, \frac{1}{16} [\slashed p \gamma_5]_{\beta \alpha} (x_1 - x_2)^2 \phi_{4 K}(u) - \frac{i}{2} [(\slashed x_1 - \slashed x_2) \gamma_5]_{\beta \alpha} \int\limits_{0}^u d v \psi_{4 K} (v) \Bigr)\,,
\nonumber
\end{eqnarray}
where $i,j,$ denote the quark colour indices, $\alpha, \beta$, are spinor indices, $f_K$ is the kaon decay constant, 
$\phi_{2K}$, \ldots, 
$\psi_{4K}$, are the kaon LCDAs of twist-2, 3, and 4, 
$\bar u = 1 - u$, and $\mu_K = m_K^2/(m_u + m_s)$ denotes the chirally enhanced parameter. 
The diagram relevant for the derivation of the LC-OPE is shown in Fig.~\ref{fig:diagrams-OPE}~\footnote{
The computation of the correlator in Eq.~\eqref{eq:Correlator-O1} actually leads also to a second diagram corresponding to an annihilation topology. However, its contribution to the dispersion relations vanishes, as expected, given the LO-QCD accuracy of our analysis.}, 
and its computation leads to the following result
\begin{equation}
F_q (p^2, q^2)\Bigl|_{\rm OPE}  = m_c \, f_K \int\limits_{0}^1 \! du \sum_{\phi}  \phi (u) \sum_{n=1}^3 \frac{c^\phi_n (u, q^2)}{\bigl[\tilde s (u, q^2) - p^2 \bigr]^n} \ln \left(\frac{m_s^2 - q^2}{\mu^2}\right)\, ,
\label{eq:Fq-OPE}
\end{equation}
where $\phi = \{ \phi_{2K} (u), \ldots, \psi_{4K} (u)\} $, and we have only retained the logarithmic term arising from the loop calculation, as this is the only relevant for the derivation of the dispersion relations.
Moreover, the function $\tilde s (u, q^2)$ in Eq.~\eqref{eq:Fq-OPE} reads
\begin{equation}
\tilde s (u, q^2) = \frac{m_c^2 - \bar u q^2 + u \bar u m_K^2}{u}\,,
\end{equation}
and note that the coefficients $c_\phi^n (u, q^2)$ have been suitably manipulated so that the dependence on $p^2$ is contained only in the denominators. Their explicit expressions can be found in the Appendix.\\
The OPE result in Eq.~\eqref{eq:Fq-OPE} can be then linked to the desired matrix element via the derivation of hadronic dispersion relations, see e.g.~Refs.~\cite{Khodjamirian:2000mi, Piscopo:2023opf} for details.
In particular, after isolating the ground $D^0$- and $K^+$-meson states in the $p^2$- and $q^2$-channels, respectively, the contribution of 
the excited states and of the continuum is approximated by means of the quark hadron duality, the latter introducing two effective threshold parameters $s_0^D$ and $s_0^K$. Finally, a Borel transformation in both channels i.e.\ $p^2 \to M^2$, $q^2 \to M^{\prime^2}$, is performed in order to suppress the contribution of the continuum. 
The final sum-rule takes then the form
\begin{equation}
    i \langle K^+ K^- |O_1^s| D^0 \rangle = \frac{e^{m_{D}^2/M^2} e^{m_{K}^2/M^{\prime 2}}}{\pi^2 f_K f_D \, m_D^2} \int\limits_{m_s^2}^{s_0^K} d s^\prime \!  \int\limits_{m_c^2}^{s_0^D} d s \, e^{-s/M^2}   e^{-s^\prime/M^{\prime 2}}\,  {\rm Im}_{s^\prime}{\rm Im}_s \bigl[ F_q (s,s^\prime) \bigr]_{\rm OPE} \,,
    \label{eq:O1s-SR}
\end{equation}
where the expression for the imaginary part of $F_q(s,s^\prime)|_{\rm OPE}$ can be easily derived using the results given e.g.\ in the Appendix of Ref.~\cite{Piscopo:2023opf}.
\section{Numerical analysis and results}
\label{sec:Result}
\renewcommand{\arraystretch}{1.4}
\begin{table}[t]
    \centering
    \begin{tabular}{|c|c|c|c|c|c|}
    \hline
    \multicolumn{6}{|c|}{Parameters of the LCDAs} \\
    \hline
    $\mu_\pi $ & $(2.50 \pm 0.30) \GeV$ &  \cite{Khodjamirian:2017fxg} &
    $\mu_K $ & $(2.49 \pm 0.26) \GeV$  & \cite{Khodjamirian:2017fxg} \\
    $a^\pi_2$ & $0.275 \pm 0.055$ & \cite{Cheng:2020vwr} &
    $a^K_1$ & $0.10 \pm 0.04$  & \cite{Chetyrkin:2007vm} \\
    $a^\pi_4$ & $0.185 \pm 0.065$ & \cite{Cheng:2020vwr} &
    $a^K_2$ & $0.25 \pm 0.15$  & \cite{Ball:2007zt} \\ 
    $f_3^\pi$ & $(0.0045 \pm 0.0015) \GeV^2$ & \cite{Ball:2007zt} &
    $f_3^K$ & $(0.0045 \pm 0.0020) \GeV^2$ & \cite{Ball:2007zt} \\
    $\omega_3^\pi$ & $-1.5 \pm 0.7$ & \cite{Ball:2007zt} &
    $\omega_3^K$ & $ -1.2 \pm 0.7$  & \cite{Ball:2007zt} \\
    $\lambda_3^\pi$ & $0$ &  --- &
    $\lambda_3^K$ & $ 1.6 \pm 0.4$  & \cite{Ball:2007zt}\\
    $\delta_\pi^2$ & $(0.18 \pm 0.06) \GeV^2$ & \cite{Ball:2007zt} &
    $\delta_K^2$ & $(0.20 \pm 0.06) \GeV^2$  & \cite{Ball:2007zt}\\
    $\omega_4^\pi$ & $0.2 \pm 0.1$ & \cite{Ball:2007zt}  &
    $\omega_4^K$ & $ 0.2 \pm 0.1$  &  \cite{Ball:2007zt}\\
    $\kappa_{4\pi}$ & $0$ & --- &
    $\kappa_{4 K}$ & $ - 0.12 \pm 0.01 $  & \cite{Ball:2007zt} \\
    \hline
    \multicolumn{6}{|c|}{Sum rule parameters} \\
    \hline
    $s_0^\pi$ & $(0.7 \pm 0.1) \GeV^2$ & \cite{Khodjamirian:2017zdu} & 
    $s_0^K$ & $(1.2 \pm 0.1) \GeV^2$  & \cite{Khodjamirian:2017zdu} \\
    $M^2_\pi$ & $(1.0 \pm 0.5) \GeV^2$ & \cite{Khodjamirian:2006st} & 
    $M^2_K$ & $(1.0 \pm 0.5) \GeV^2$ & \cite{Khodjamirian:2006st} \\
    $s_0^D$ & $8.2^{+1.4}_{-0.6} \GeV^2$ & \eqref{eq:mDsq-SR} & 
    $M^2_D$ & $(4.5 \pm 1.0) \GeV^2$ & \cite{Khodjamirian:2009ys} \\
    \hline 
    \multicolumn{6}{|c|}{CKM parameters} \\
    \hline
    $|V_{us}|$ & $0.22500^{+0.00024}_{-0.00021}$ & \cite{Charles:2004jd} &
    $\frac{|V_{ub}|}{|V_{cb}|}$ &  $0.08848^{+0.00224}_{-0.00219}$ &
    \cite{Charles:2004jd} \\
    $|V_{cb}|$ &  $0.04145^{+0.00035}_{-0.00061}$ & \cite{Charles:2004jd} &
    $\delta$ & $\left(65.5^{+1.3}_{-1.2}\right)^\circ$ & \cite{Charles:2004jd} \\
    \hline 
    \multicolumn{6}{|c|}{Other parameters} \\
    \hline
    $m_{\pi^\pm}$ & $0.13957 \GeV$ & \cite{Workman:2022ynf} &
    $m_{K^\pm}$   & $0.493677 \GeV$ & \cite{Workman:2022ynf} \\
    $f_\pi $ & $(0.1302 \pm 0.0008) \GeV$ & \cite{FLAG} &
    $f_K$ & $(0.1557 \pm 0.0003) \GeV$  & \cite{FLAG} \\
    \hline
    $\overline{m}_c$ & 
    $(1.27 \pm 0.02) \GeV$ & \cite{Workman:2022ynf} &
    $\overline{m}_s$ & $(0.0934^{+0.0086}_{-0.0034}) \GeV $ & \cite{Workman:2022ynf} \\
    $\alpha_s (m_Z) $ & $0.1179 \pm 0.0009 $ & \cite{Workman:2022ynf} &
    $m_{D^0} $ & $1.86484 \GeV$ & \cite{Workman:2022ynf} \\
    $f_D$ & $(0.2120 \pm  0.0007) \GeV$ & \cite{FLAG} &
    $\tau (D^0)$ & $(0.4013 \pm 0.0010) \, {\rm ps}$ & \cite{Workman:2022ynf} \\ 
    \hline
    \end{tabular}
    \caption{Values of the parameters used in the numerical analysis. All numbers quoted correspond to the  non-perturbative parameters evaluated at the renormalistation scale $\mu = 1 \GeV$, apart from $\mu_\pi$ and $\mu_K$, whose values are given at $\mu = 2 \GeV$.}
    \label{tab:input}
\end{table}
In this section we discuss the choice of the input parameters and present our results for the leading contribution to the amplitudes ${\cal A_{\pi\pi}}$, ${\cal A}_{K K}$, in Eqs.~\eqref{eq:Apipi-appr}, \eqref{eq:AKK-appr}.
For convenience, all the values used in the numerical analysis are also summarised in Table~\ref{tab:input}.\\
For the pion and kaon LCDAs up to twist-4, we use the same expressions as in the corresponding LCSR studies of the $D \to \pi$ and $D \to K$ form factors~\cite{Khodjamirian:2009ys}. In particular, the twist-2 LCDAs are obtained in the form of a truncated expansion in the Gegenbauer polynomials $C_n^{(3/2)} (u - \bar u)$ with  corresponding coefficients $a^\pi_n$ and $a^K_n$, respectively. For the pion LCDAs, the odd Gegenbauer coefficients vanish, the values of $a_2^\pi$ and $a_4^\pi$ are taken from the recent study~\cite{Cheng:2020vwr}, while we neglect $a_{n>4}^\pi$.
For the kaon LCDAs, we take the value of $a_1^K$ from Ref.~\cite{Chetyrkin:2007vm}, of $a_2^K$ from Ref.~\cite{Ball:2007zt}, and neglect $a_{n>2}^K$.
The value of the chirally-enhanced parameters $\mu_\pi$ and $\mu_K$, entering the twist-3 LCDAs, are obtained by employing well-known relations in chiral perturbation theory~\cite{Leutwyler:1996qg}, and can be found e.g.\ in Ref.~\cite{Khodjamirian:2017fxg}. 
The remaining parameters entering the twist-3 and twist-4 LCDAs are taken to be the same as in the study~\cite{Khodjamirian:2009ys}, corresponding to the values obtained in Ref.~\cite{Ball:2007zt}.
As for the renormalisation scale~$\mu$, following Ref.~\cite{Khodjamirian:2009ys}, we fix the central value to $\mu = 1.5 \GeV \sim \sqrt{m_D^2 - m_c^2}$, and vary this in the interval $1 \GeV \leq \mu \leq 2 \GeV$.
The scale dependence at one-loop accuracy for the LCDAs parameters is taken from Ref.~\cite{Ball:2007zt}.
The running of the strong coupling, as well as of the quark masses
in the $\overline{\rm MS}$-scheme, is implemented using the Mathematica package $\mathtt{ RunDec}$~\cite{Herren:2017osy}, fixing the values of $\alpha_s (m_Z)$, $\overline{m}_c (\overline m_c)$, and $\overline{m}_s (2 \GeV)$ as given in Ref.~\cite{Workman:2022ynf}.\\
Meson decay constants
are determined precisely from Lattice QCD calculations, and we take their values from Ref.~\cite{FLAG}.
Meson masses, also known very precisely, are taken from the PDG~\cite{Workman:2022ynf}.
Finally, to predict branching fractions and $\Delta a^{\rm dir}_{\rm CP}$, we need in addition to fix the value of the Wilson coefficients and the CKM parameters. Leading-order results for $C_1 (\mu_1)$ and $C_2 (\mu_1)$ are implemented using the expressions given in Ref.~\cite{Buchalla:1995vs}, see also Table~3 of Ref.~\cite{King:2021xqp}, and we vary the renormalisation scale in the interval $\mu_1 = (1.5 \pm 0.5) \GeV$. 
As for the CKM matrix elements, we use the standard parametrisation, taking the values of 
$|V_{us}|$, $|V_{cb}|$, $|V_{ub}/V_{cb}|$, and~$\delta$, from the CKMfitter~\cite{Charles:2004jd}
(similar values can be obtained from the UTFit~\cite{UTfit:2006vpt}).\\
Concerning the choice of the sum rule parameters, we mostly follow Ref.~\cite{Khodjamirian:2017zdu}.
Thus, for the Borel parameters we adopt the same interval $M^{\prime 2} = (1 \pm 0.5) \GeV^2$ for both the pion and kaon channels, while for the respective threshold  continuum parameters we use $s^\pi_0 = (0.7 \pm 0.1) \GeV^2$ and $s^K_0 = (1.2 \pm 0.1) \GeV^2$~\cite{Braun:1999uj, Bijnens:2002mg}.
The Borel parameter for the $D$-meson channel is chosen to be in the interval $M^2 = (4.5 \pm 1.0) \GeV^2$, following Ref.~\cite{Khodjamirian:2009ys},
while we fix the threshold continuum parameter $s_0^D$ by differentiating with respect to $1/M^2$ both sides of the sum-rule in the case of pion final state, cf.~Eq.~\eqref{eq:O1s-SR}, obtaining the following relation
\begin{equation}
[m_{D^0}^2]_{\rm LCSR} = \frac{\displaystyle\int_{m_c^2}^{s_0^D} d s \, s\,  e^{-s/M^2} \int_{0}^{s_0^\pi} d s^\prime e^{-s^\prime/M^{\prime 2}} \, {\rm Im}_{s^\prime}{\rm Im}_s \bigl[ F_q (s,s^\prime) \bigr]_{\rm OPE}}{\displaystyle \int_{m_c^2}^{s_0^D} d s \, e^{-s/M^2} \int_{0}^{s_0^\pi} d s^\prime e^{-s^\prime/M^{\prime 2}} \, {\rm Im}_{s^\prime}{\rm Im}_s \bigl[ F_q (s,s^\prime) \bigr]_{\rm OPE}}\,.  
\label{eq:mDsq-SR}
\end{equation}
The sum-rule result for the $D^0$-meson mass squared is then fitted to its experimental value, leading to  $s_0^D \approx 8.2 \GeV^2$ in correspondence of the central values of the remaining input parameters. The uncertainty on $s_0^D$, shown in Table~\ref{tab:input}, is derived varying $M^2$ within its error range.\\
Using the values of the input parameters as described above, we obtain the following LCSR predictions for the tree-level matrix elements of the four modes analysed~\footnote{The operators introduced in Eqs.~\eqref{eq:MEO1-K-pi}, \eqref{eq:MEO1-pi-K}, read $O_1^{q_1q_2} = (\bar q_1^i \Gamma_\mu c^i) (\bar u^j \Gamma^\mu q_2^j)$, with $q_i = d,s$.}
\begin{eqnarray}
\hspace*{-5mm}
\langle K^+ K^- | O_1^{s} | D^0 \rangle\Bigl|_{\rm LCSR} 
& = & i \left(0.413^{+0.069}_{-0.111} \pm 0.165\right)\GeV^3
=i \, 0.413^{+0.179}_{-0.199} \GeV^3,
\label{eq:MEO1-K-K} \\
\hspace*{-5mm}
\langle \pi^+ \pi^- | O_1^{d} | D^0 \rangle\Bigl|_{\rm LCSR} 
& = & i \left(0.236^{+0.047}_{-0.073} \pm 0.094\right)\GeV^3
= i \,0.236^{+0.105}_{-0.119} \GeV^3 \,,
\label{eq:MEO1-pi-pi} \\
\hspace*{-5mm}
\langle \pi^+ K^- | O_1^{sd} | D^0 \rangle\Bigl|_{\rm LCSR} 
& = & i \left(0.261^{+0.049}_{-0.079} \pm 0.103\right)\GeV^3
= i \, 0.259^{+0.116}_{-0.131} \GeV^3 \,,
\label{eq:MEO1-K-pi} \\
\hspace*{-5mm}
\langle K^+ \pi^-  | O_1^{ds} | D^0 \rangle\Bigl|_{\rm LCSR} 
& = & i \left(0.391^{+0.068}_{-0.105} \pm 0.156\right)\GeV^3
= i \, 0.390^{+0.170}_{-0.189} \GeV^3 \,,
\label{eq:MEO1-pi-K}
\end{eqnarray}
where the first uncertainties are due to variation of all the input parameters and of the renormalisation scale, and the second account for missing  higher-order QCD effects, which we conservatively estimate within the $40 \%$ range.\\
With the LCSR results in Eq.~\eqref{eq:MEO1-K-K} - \eqref{eq:MEO1-pi-K}, and the expressions derived in Section~\ref{sec:theory-framework} for the amplitude and branching fraction, cf.~Eqs.~\eqref{eq:BR-D0-to-pi-pi}, \eqref{eq:Apipi-appr}, we arrive at the following values for the decays considered
\begin{align}
{\cal B} (D^0 \to K^+ K^-)\bigl|_{\rm LCSR}  & =  \left(3.67^{+3.90}_{-2.69}\right) \times 10^{-3}\,,
\label{eq:BR-D0-to-K-K-LCSR} \\[2mm]
{\cal B} (D^0 \to \, \pi^+ \, \pi^-)\bigl|_{\rm LCSR}  & =  \left(1.40^{+1.53}_{-1.06}\right) \times 10^{-3}\,,
\label{eq:BR-D0-to-pi-pi-LCSR} \\[2mm]
{\cal B} (D^0 \to \pi^+ K^-)\bigl|_{\rm LCSR}   & =  \left(2.99^{+3.26}_{-2.26}\right) \times 10^{-2}\,,
\label{eq:BR-D0-to-K-pi-LCSR}  \\[2mm]
{\cal B} (D^0 \to K^+ \pi^-)\bigl|_{\rm LCSR}  & =  \left(1.80^{+1.93}_{-1.33}\right) \times 10^{-4}\,,
\label{eq:BR-D0-to-pi-K-LCSR}
\end{align}
in excellent agreement with the experimental data in 
Eqs.~\eqref{eq:Br_exp-KK} - \eqref{eq:Br_exp-piK}. The quoted uncertainties are sizable and largely follow from our very conservative treatment of missing corrections.
Again, these results do not indicate the presence of unexpectedly big sub-leading effects. \\
Next, we discuss our estimate for $|\Delta a_{\rm CP}^{\rm dir}|$. With the results for ${\cal A}_{KK}$, ${\cal A}_{\pi \pi}$, obtained using Eqs.~\eqref{eq:MEO1-K-K}, \eqref{eq:MEO1-pi-pi}, and implementing the expressions for the penguin matrix elements ${\cal P}_{KK}$, ${\cal P}_{\pi \pi}$, from Ref.~\cite{Khodjamirian:2017zdu}, in order to account for correlations
originating from the use of the same theoretical framework and input parameters, we obtain the following values for the ratios entering Eqs.~\eqref{ew:a_pipi_appr}, \eqref{ew:a_KK_appr}, that is
\begin{equation}
\left|\frac{{\cal P}_{KK}}{{\cal A}_{KK}} \right|_{\rm LCSR}
 =  0.066^{+0.031}_{-0.029}\,, 
\qquad
\left|\frac{{\cal P}_{\pi \pi}}{{\cal A}_{\pi \pi}} \right|_{\rm LCSR}
 =  0.089^{+0.042}_{-0.037}\,, 
\label{eq:P-to-A-LCSR} 
\end{equation}
in perfect agreement with the results of Ref.~\cite{Khodjamirian:2017zdu}. Note that we have again added a conservative $40 \%$ uncertainty, to account for missing higher-order QCD contributions, higher-twist effects, and corrections of the order $O(s_0^{K,\pi}/m_D^2)$ not included in Ref.~\cite{Khodjamirian:2017zdu}.  
Using the results in Eq.~\eqref{eq:P-to-A-LCSR}, and allowing for arbitrary strong phase differences, that is varying both $\sin \phi_{\pi \pi}$ and $\sin \phi_{K K}$ from $-1$ to $1$ in Eq.~\eqref{eq:Delta-ACP}, we obtain the following upper bound for $|\Delta a_{\rm CP}^{\rm dir}|$, namely
\begin{equation}
|\Delta a_{\rm CP}^{\rm dir}|_{\rm LCSR} \leq 2.4 \times 10^{-4} \,,
\label{eq:upperbound}
\end{equation}
which is about 6 times smaller than the current central experimental value in Eq.~\eqref{eq:DeltaACP_exp}.\\
Before concluding, it is worth emphasising that we also computed the tree-level matrix elements in Eqs.~\eqref{eq:Apipi-appr}, \eqref{eq:AKK-appr}, using a LCSR with $D$-meson LCDAs. In the latter case, which largely follows the calculation performed in the recent study of non-leptonic $B$-meson decays~\cite{Piscopo:2023opf}, the parameters of the $D$-meson LCDAs were taken from the corresponding ones for the $B$-meson assuming heavy quark flavour symmetry. Interestingly, including contributions with two-particle $D$-meson LCDAs up to twist-3, we obtained central values in agreement with the results in Eqs.~\eqref{eq:BR-D0-to-K-K-LCSR}, \eqref{eq:BR-D0-to-pi-pi-LCSR}. On the other hand, the respective uncertainties were also found to be huge, reflecting the poor precision with which the $D$-meson LCDAs are known.

\section{Conclusion and outlook}
\label{sec:Conclusion}
In this paper we have employed the framework of LCSR to determine the SM predictions for several non-leptonic $D^0$ decays. Focusing on the computation of the leading contribution to the decay amplitude, i.e.\ considering only the color-allowed tree-level topology at leading order in QCD and including two-particle pion/kaon LCDAs up to twist-4, we find an astonishing agreement with the experimental values of the branching ratios for the modes 
$D^0 \to \pi^+ K^-$, $D^0 \to K^+ K^-$, $D^0 \to \pi^+ \pi^- $ 
and $D^0 \to K^+ \pi^-$. 
Our results for the decay amplitudes, combined with the expressions for the penguin diagrams obtained within LCSR in Ref.~\cite{Khodjamirian:2017zdu}, lead to a bound for $|\Delta a_{\rm CP}^{\rm dir}|$ which is considerably lower than the current experimental determination.
\\
To obtain a more profound statement, 
we plan to extend our theoretical study to include the following improvements:
\begin{enumerate}
\item 
Calculate the contribution of condensate and soft-gluon effects, i.e.\ three-particle LCDAs. 
Furthermore, higher-twist corrections
could be investigated.

\item Calculate higher order perturbative QCD corrections to the sum-rule result. 
This will also allow to extend our approach to other topologies than the color-allowed tree-level one, like annihilation and penguin diagrams, and will have a crucial impact on the theoretical determination of the strong phases.

\item Extend our study to decays that are governed by   color-suppressed tree-level  topologies.

\end{enumerate}
In light of the surprising agreement of our results - and in particular of the nQCDf estimates - with the experimental branching ratios,
we consider it to be valuable to contemplate a scenario in which the above listed future improvements to the sum-rule predictions will not change considerably  the current picture.
In such a scenario we could conclude:
\begin{itemize}
\item[$\diamond$] The framework of LCSR can be successfully employed to predict the branching ratios for the decays $D^0 \to \pi^+ \pi^-, \, 
 K^+ \pi^-, \,  \pi^+ K^-, \,  K^+ K^-$.
The color-allowed tree-level diagrams give the dominant contribution to the branching fractions and the remaining topologies only lead to smaller corrections.
The current large uncertainties in theoretical predictions presented in Eqs.~\eqref{eq:BR-D0-to-K-K-LCSR} - \eqref{eq:BR-D0-to-pi-K-LCSR} can be systematically reduced including the improvements listed above.
\item[$\diamond$] The size of SU(3)$_f$ breaking effects turns out to be very large and it is well accommodated by the values of the decay constants and form factors. Comparing the central values of Eq.~\eqref{eq:MEO1-K-K} and Eq.~\eqref{eq:MEO1-pi-pi} we find SU(3)$_f$ breaking effects of the order of $75 \%$ at the amplitude level, thus questioning the applicability of this symmetry for $D$-meson decays. In this respect, we would like to note that studies like Refs.~\cite{Hiller:2012xm, Hiller:2013awa, Muller:2015lua} only set a lower limit of $30 \%$ on the possible size of the SU(3)$_f$ breaking, while allowing also 
much larger values, in consistency with our finding.

\item[$\diamond$] We find an upper bound on $|\Delta a_{\rm CP}^{\rm dir}|$, given in Eq.~\eqref{eq:upperbound}, which is about a factor of 6 smaller than the current experimental average given in Eq.~\eqref{eq:DeltaACP_exp}. If the experimental numbers will stay, in particular the central values for the individual CP asymmetries given in Eqs.~\eqref{eq:DeltaACP_KK_exp}, \eqref{eq:DeltaACP_pipi_exp}, then this could be a first glimpse of physics beyond the SM.

\end{itemize}

\section*{Acknowledgements}
The authors would like to thank Alexander Khodjamirian and Thomas Mannel for valuable comments. MLP wishes to thank Vladyslav Shtabovenko for helpful discussions.
The work of MLP was funded by the Deutsche Forschungsgemeinschaft (DFG, German
Research Foundation) - project number 500314741.
The work of AR is supported by the DFG, under grant 396 021 762 - TRR 257 “Particle Physics Phenomenology after the Higgs Discovery”.

\section*{Note added:}
While this work was being completed, Ref.~\cite{Gavrilova:2023fzy} appeared on the arXiv. Assuming the validity of the SM for the description of $D^0$ decays, the authors find experimental evidence for large penguin/rescattering effects. These findings do not change the conclusions of our paper, in which first steps towards a fully QCD based calculation indicate that the current experimental value for $\Delta a^{\rm dir}_{\rm CP}$ cannot be reproduced in the SM.

\appendix
\section{Expressions of the OPE coefficients}
Here, we present non-vanishing coefficients $c_n^\phi (u, q^2)$,
introduced in Eq.~\eqref{eq:Fq-OPE}.
\begin{eqnarray}
c^{\phi_{2K}}_1 & = & 
-\frac{m_c \left(m_s^2-q^2\right){}^2}{8 \pi ^2 q^6 u^2} \Bigl[m_c^2 \left(2 m_s^2+q^2\right) + q^2 \left(q^2 (2 u-1)-u^2 m_K^2\right) 
\nonumber \\
& & + m_s^2 \left(q^2 (u-2)-2 u^2 m_K^2\right) \Bigr]\, , \\[2mm]
c^{\phi_{3K}^p}_1 & = &
-\frac{\mu_K \left(m_s^2-q^2\right){}^2}{8 \pi ^2 q^6 u} \Bigl[m_c^2 \left(2 m_s^2 + q^2 \right) + q^2 \left(-u^2 m_K^2+2 q^2 u - q^2 \right) 
\nonumber \\
& & + m_s^2 \left(-2 u^2 m_K^2+q^2 u+q^2\right) \Bigr]\,, \\[2mm]
c^{\phi_{3K}^\sigma}_1 & = & 
-\frac{\mu_K \left(m_s^2-q^2\right){}^2}{48 \pi ^2 q^6 u^2} \Bigl[m_c^2 \left(2 m_s^2+q^2\right) +q^2 \left(-u^2 m_K^2+4 q^2 u -3 q^2\right) 
\nonumber \\
& & + m_s^2 \left(-2 u^2 m_K^2+2 q^2 u-3 q^2\right) \Bigr]\,, \\[2mm]
c^{\phi_{3K}^\sigma}_2 & = &
-\frac{\mu_K \left(m_s^2-q^2\right){}^2}{48 \pi ^2 q^6 u^3} \Bigl[m_c^4 \left(2m_s^2+q^2\right) 
\nonumber \\
& & + m_c^2 \left(m_s^2 \left(q^2 (u-3)-4 u^2 m_K^2\right)+2 q^2 u \left(q^2-u m_K^2\right)\right)
\\
& & +\left(q^2-u^2 m_K^2\right) \left(m_s^2 \left(-2 u^2 m_K^2+ (u+1)q^2\right)+q^2 \left(-u^2 m_K^2+ (2 u - 1) q^2 \right)\right)\Bigr]\,, 
\nonumber \\[2mm]
c^{\phi_{4K}}_2 & = & -\frac{m_c^3 \left(m_s^2-q^2\right){}^2 \left(2 m_s^2+q^2\right)}{16 \pi ^2 q^6 u^3} \, , \\[2mm]
c^{\phi_{4K}}_3 & = &
\frac{m_c^3 \left(m_s^2-q^2\right){}^2}{16 \pi ^2 q^6 u^4} \Bigl[m_c^2 \left(2 m_s^2+q^2\right)+q^2 \left(q^2 (2 u-1)-u^2 m_K^2\right) 
\nonumber \\
& & + m_s^2 \left(q^2 (u-2)-2 u^2 m_K^2\right) \Bigr]\,, \\[2mm]
c^{\psi_{4K}}_1 & = &
-\frac{m_c \left(m_s^2-q^2\right){}^2 \left(2 m_s^2+q^2\right)}{8 \pi ^2 q^6 u}\,, \\[2mm]
c^{\psi_{4K}}_2 & = &
\frac{m_c \left(m_s^2-q^2\right){}^2}{8 \pi ^2 q^6 u^2} \Bigl[m_c^2 \left(2 m_s^2+q^2\right)+q^2 \left(-u^2 m_K^2+2 q^2 u-q^2\right) 
\nonumber \\ 
& & +m_s^2 \left(-2
   u^2 m_K^2+q^2 u+q^2\right)\Bigr]\,.
\end{eqnarray}

\bibliographystyle{JHEP}
\bibliography{References}

\providecommand{\href}[2]{#2}\begingroup\raggedright\begin{thebibliography}{10}

\bibitem{Lenz:2020awd}
A.~Lenz and G.~Wilkinson, {\it {Mixing and CP Violation in the Charm System}},
  {\em Ann. Rev. Nucl. Part. Sci.} {\bf 71} (2021) 59--85,
  [\href{http://arxiv.org/abs/2011.04443}{{\tt arXiv:2011.04443}}].

\bibitem{King:2021xqp}
D.~King, A.~Lenz, M.~L. Piscopo, T.~Rauh, A.~V. Rusov, and C.~Vlahos, {\it
  {Revisiting inclusive decay widths of charmed mesons}},  {\em JHEP} {\bf 08}
  (2022) 241, [\href{http://arxiv.org/abs/2109.13219}{{\tt arXiv:2109.13219}}].

\bibitem{Gratrex:2022xpm}
J.~Gratrex, B.~Meli\'c, and I.~Ni\v{s}and\v{z}i\'c, {\it {Lifetimes of singly
  charmed hadrons}},  {\em JHEP} {\bf 07} (2022) 058,
  [\href{http://arxiv.org/abs/2204.11935}{{\tt arXiv:2204.11935}}].

\bibitem{Cheng:2023jpz}
H.-Y. Cheng and C.-W. Liu, {\it {Study of singly heavy baryon lifetimes}},
  {\em JHEP} {\bf 07} (2023) 114, [\href{http://arxiv.org/abs/2305.00665}{{\tt
  arXiv:2305.00665}}].

\bibitem{HFLAV:2022esi}
{\bf HFLAV} Collaboration, Y.~S. Amhis et~al., {\it {Averages of b-hadron,
  c-hadron, and \ensuremath{\tau}-lepton properties as of 2021}},  {\em Phys.
  Rev. D} {\bf 107} (2023), no.~5 052008,
  [\href{http://arxiv.org/abs/2206.07501}{{\tt arXiv:2206.07501}}].

\bibitem{Beneke:2002rj}
M.~Beneke, G.~Buchalla, C.~Greub, A.~Lenz, and U.~Nierste, {\it {The $B^+
  -B^0_d$ Lifetime Difference Beyond Leading Logarithms}},  {\em Nucl. Phys. B}
  {\bf 639} (2002) 389--407, [\href{http://arxiv.org/abs/hep-ph/0202106}{{\tt
  hep-ph/0202106}}].

\bibitem{Franco:2002fc}
E.~Franco, V.~Lubicz, F.~Mescia, and C.~Tarantino, {\it {Lifetime ratios of
  beauty hadrons at the next-to-leading order in QCD}},  {\em Nucl. Phys. B}
  {\bf 633} (2002) 212--236, [\href{http://arxiv.org/abs/hep-ph/0203089}{{\tt
  hep-ph/0203089}}].

\bibitem{Lenz:2013aua}
A.~Lenz and T.~Rauh, {\it {D-meson lifetimes within the heavy quark
  expansion}},  {\em Phys. Rev. D} {\bf 88} (2013) 034004,
  [\href{http://arxiv.org/abs/1305.3588}{{\tt arXiv:1305.3588}}].

\bibitem{Lenz:2020oce}
A.~Lenz, M.~L. Piscopo, and A.~V. Rusov, {\it {Contribution of the Darwin
  operator to non-leptonic decays of heavy quarks}},  {\em JHEP} {\bf 12}
  (2020) 199, [\href{http://arxiv.org/abs/2004.09527}{{\tt arXiv:2004.09527}}].

\bibitem{Mannel:2020fts}
T.~Mannel, D.~Moreno, and A.~Pivovarov, {\it {Heavy quark expansion for heavy
  hadron lifetimes: completing the $ 1/{m}_b^3 $ corrections}},  {\em JHEP}
  {\bf 08} (2020) 089, [\href{http://arxiv.org/abs/2004.09485}{{\tt
  arXiv:2004.09485}}].

\bibitem{Moreno:2020rmk}
D.~Moreno, {\it {Completing $1/m_b^3$ corrections to non-leptonic
  bottom-to-up-quark decays}},  {\em JHEP} {\bf 01} (2021) 051,
  [\href{http://arxiv.org/abs/2009.08756}{{\tt arXiv:2009.08756}}].

\bibitem{Kirk:2017juj}
M.~Kirk, A.~Lenz, and T.~Rauh, {\it {Dimension-six matrix elements for meson
  mixing and lifetimes from sum rules}},  {\em JHEP} {\bf 12} (2017) 068,
  [\href{http://arxiv.org/abs/1711.02100}{{\tt arXiv:1711.02100}}]. [Erratum:
  JHEP 06, 162 (2020)].

\bibitem{King:2021jsq}
D.~King, A.~Lenz, and T.~Rauh, {\it {SU(3) breaking effects in B and D meson
  lifetimes}},  {\em JHEP} {\bf 06} (2022) 134,
  [\href{http://arxiv.org/abs/2112.03691}{{\tt arXiv:2112.03691}}].

\bibitem{Black:2023vju}
M.~Black, R.~Harlander, F.~Lange, A.~Rago, A.~Shindler, and O.~Witzel, {\it
  {Using Gradient Flow to Renormalise Matrix Elements for Meson Mixing and
  Lifetimes}},  in {\em {40th International Symposium on Lattice Field
  Theory}}, 10, 2023.
\newblock \href{http://arxiv.org/abs/2310.18059}{{\tt arXiv:2310.18059}}.

\bibitem{Glashow:1970gm}
S.~L. Glashow, J.~Iliopoulos, and L.~Maiani, {\it {Weak Interactions with
  Lepton-Hadron Symmetry}},  {\em Phys. Rev. D} {\bf 2} (1970) 1285--1292.

\bibitem{Falk:2001hx}
A.~F. Falk, Y.~Grossman, Z.~Ligeti, and A.~A. Petrov, {\it {SU(3) breaking and
  $D^0 - \bar{D}^0$ mixing}},  {\em Phys. Rev. D} {\bf 65} (2002) 054034,
  [\href{http://arxiv.org/abs/hep-ph/0110317}{{\tt hep-ph/0110317}}].

\bibitem{Falk:2004wg}
A.~F. Falk, Y.~Grossman, Z.~Ligeti, Y.~Nir, and A.~A. Petrov, {\it {The $D^0 -
  \bar{D}^0$ mass difference from a dispersion relation}},  {\em Phys. Rev. D}
  {\bf 69} (2004) 114021, [\href{http://arxiv.org/abs/hep-ph/0402204}{{\tt
  hep-ph/0402204}}].

\bibitem{Bobrowski:2010xg}
M.~Bobrowski, A.~Lenz, J.~Riedl, and J.~Rohrwild, {\it {How Large Can the SM
  Contribution to CP Violation in $D^0-\bar D^0$ Mixing Be?}},  {\em JHEP} {\bf
  03} (2010) 009, [\href{http://arxiv.org/abs/1002.4794}{{\tt
  arXiv:1002.4794}}].

\bibitem{Lenz:2020efu}
A.~Lenz, M.~L. Piscopo, and C.~Vlahos, {\it {Renormalization scale setting for
  D-meson mixing}},  {\em Phys. Rev. D} {\bf 102} (2020), no.~9 093002,
  [\href{http://arxiv.org/abs/2007.03022}{{\tt arXiv:2007.03022}}].

\bibitem{LHCb:2011osy}
{\bf LHCb} Collaboration, R.~Aaij et~al., {\it {Evidence for CP violation in
  time-integrated $D^0 \to h^-h^+$ decay rates}},  {\em Phys. Rev. Lett.} {\bf
  108} (2012) 111602, [\href{http://arxiv.org/abs/1112.0938}{{\tt
  arXiv:1112.0938}}].

\bibitem{LHCb:2019hro}
{\bf LHCb} Collaboration, R.~Aaij et~al., {\it {Observation of CP Violation in
  Charm Decays}},  {\em Phys. Rev. Lett.} {\bf 122} (2019), no.~21 211803,
  [\href{http://arxiv.org/abs/1903.08726}{{\tt arXiv:1903.08726}}].

\bibitem{Lenz:2013pwa}
A.~Lenz, {\it {What did we learn in theory from the $\Delta A_{CP}$-saga?}},
  in {\em {6th International Workshop on Charm Physics}}, 11, 2013.
\newblock \href{http://arxiv.org/abs/1311.6447}{{\tt arXiv:1311.6447}}.

\bibitem{Chala:2019fdb}
M.~Chala, A.~Lenz, A.~V. Rusov, and J.~Scholtz, {\it {$\Delta A_{CP}$ within
  the Standard Model and beyond}},  {\em JHEP} {\bf 07} (2019) 161,
  [\href{http://arxiv.org/abs/1903.10490}{{\tt arXiv:1903.10490}}].

\bibitem{LHCb:2022lry}
{\bf LHCb} Collaboration, R.~Aaij et~al., {\it {Measurement of the
  Time-Integrated CP Asymmetry in $D^0 \to K^- K^+$ Decays}},  {\em Phys. Rev.
  Lett.} {\bf 131} (2023), no.~9 091802,
  [\href{http://arxiv.org/abs/2209.03179}{{\tt arXiv:2209.03179}}].

\bibitem{Grossman:2006jg}
Y.~Grossman, A.~L. Kagan, and Y.~Nir, {\it {New physics and CP violation in
  singly Cabibbo suppressed D decays}},  {\em Phys. Rev. D} {\bf 75} (2007)
  036008, [\href{http://arxiv.org/abs/hep-ph/0609178}{{\tt hep-ph/0609178}}].

\bibitem{Khodjamirian:2017zdu}
A.~Khodjamirian and A.~A. Petrov, {\it {Direct CP asymmetry in $D\to
  \pi^-\pi^+$ and $D\to K^-K^+$ in QCD-based approach}},  {\em Phys. Lett. B}
  {\bf 774} (2017) 235--242, [\href{http://arxiv.org/abs/1706.07780}{{\tt
  arXiv:1706.07780}}].

\bibitem{Balitsky:1989ry}
I.~I. Balitsky, V.~M. Braun, and A.~V. Kolesnichenko, {\it {Radiative Decay
  $\Sigma^+ \to p \gamma$ in Quantum Chromodynamics}},  {\em Nucl. Phys. B}
  {\bf 312} (1989) 509--550.

\bibitem{Pich:2023kim}
A.~Pich, E.~Solomonidi, and L.~Vale~Silva, {\it {Final-state interactions in
  the CP asymmetries of charm-meson two-body decays}},  {\em Phys. Rev. D} {\bf
  108} (2023), no.~3 036026, [\href{http://arxiv.org/abs/2305.11951}{{\tt
  arXiv:2305.11951}}].

\bibitem{Dery:2019ysp}
A.~Dery and Y.~Nir, {\it {Implications of the LHCb discovery of CP violation in
  charm decays}},  {\em JHEP} {\bf 12} (2019) 104,
  [\href{http://arxiv.org/abs/1909.11242}{{\tt arXiv:1909.11242}}].

\bibitem{Calibbi:2019bay}
L.~Calibbi, T.~Li, Y.~Li, and B.~Zhu, {\it {Simple model for large CP violation
  in charm decays, $B$-physics anomalies, muon $g-2$ and dark matter}},  {\em
  JHEP} {\bf 10} (2020) 070, [\href{http://arxiv.org/abs/1912.02676}{{\tt
  arXiv:1912.02676}}].

\bibitem{Bause:2020obd}
R.~Bause, H.~Gisbert, M.~Golz, and G.~Hiller, {\it {Exploiting $CP$-asymmetries
  in rare charm decays}},  {\em Phys. Rev. D} {\bf 101} (2020), no.~11 115006,
  [\href{http://arxiv.org/abs/2004.01206}{{\tt arXiv:2004.01206}}].

\bibitem{Grossman:2019xcj}
Y.~Grossman and S.~Schacht, {\it {The emergence of the $\Delta U=0$ rule in
  charm physics}},  {\em JHEP} {\bf 07} (2019) 020,
  [\href{http://arxiv.org/abs/1903.10952}{{\tt arXiv:1903.10952}}].

\bibitem{Schacht:2021jaz}
S.~Schacht and A.~Soni, {\it {Enhancement of charm CP violation due to nearby
  resonances}},  {\em Phys. Lett. B} {\bf 825} (2022) 136855,
  [\href{http://arxiv.org/abs/2110.07619}{{\tt arXiv:2110.07619}}].

\bibitem{Bediaga:2022sxw}
I.~Bediaga, T.~Frederico, and P.~C. Magalh\~aes, {\it {Enhanced Charm CP
  Asymmetries from Final State Interactions}},  {\em Phys. Rev. Lett.} {\bf
  131} (2023), no.~5 051802, [\href{http://arxiv.org/abs/2203.04056}{{\tt
  arXiv:2203.04056}}].

\bibitem{Li:2019hho}
H.-N. Li, C.-D. L\"u, and F.-S. Yu, {\it {Implications on the first observation
  of charm CPV at LHCb}},  \href{http://arxiv.org/abs/1903.10638}{{\tt
  arXiv:1903.10638}}.

\bibitem{Cheng:2019ggx}
H.-Y. Cheng and C.-W. Chiang, {\it {Revisiting CP violation in $D\to P\!P$ and
  $V\!P$ decays}},  {\em Phys. Rev. D} {\bf 100} (2019), no.~9 093002,
  [\href{http://arxiv.org/abs/1909.03063}{{\tt arXiv:1909.03063}}].

\bibitem{Wang:2020gmn}
D.~Wang, C.-P. Jia, and F.-S. Yu, {\it {A self-consistent framework of
  topological amplitude and its $SU(N)$ decomposition}},  {\em JHEP} {\bf 21}
  (2020) 126, [\href{http://arxiv.org/abs/2001.09460}{{\tt arXiv:2001.09460}}].

\bibitem{Schacht:2022kuj}
S.~Schacht, {\it {A U-spin anomaly in charm CP violation}},  {\em JHEP} {\bf
  03} (2023) 205, [\href{http://arxiv.org/abs/2207.08539}{{\tt
  arXiv:2207.08539}}].

\bibitem{Bause:2022jes}
R.~Bause, H.~Gisbert, G.~Hiller, T.~H\"ohne, D.~F. Litim, and T.~Steudtner,
  {\it {U-spin-CP anomaly in charm}},  {\em Phys. Rev. D} {\bf 108} (2023),
  no.~3 035005, [\href{http://arxiv.org/abs/2210.16330}{{\tt
  arXiv:2210.16330}}].

\bibitem{ParticleDataGroup:2022pth}
{\bf Particle Data Group} Collaboration, R.~L. Workman et~al., {\it {Review of
  Particle Physics}},  {\em PTEP} {\bf 2022} (2022) 083C01.

\bibitem{Buchalla:1995vs}
G.~Buchalla, A.~J. Buras, and M.~E. Lautenbacher, {\it {Weak decays beyond
  leading logarithms}},  {\em Rev. Mod. Phys.} {\bf 68} (1996) 1125--1144,
  [\href{http://arxiv.org/abs/hep-ph/9512380}{{\tt hep-ph/9512380}}].

\bibitem{Khodjamirian:2000mi}
A.~Khodjamirian, {\it {$B \to \pi \pi$ decay in QCD}},  {\em Nucl. Phys. B}
  {\bf 605} (2001) 558--578, [\href{http://arxiv.org/abs/hep-ph/0012271}{{\tt
  hep-ph/0012271}}].

\bibitem{Khodjamirian:2003eq}
A.~Khodjamirian, T.~Mannel, and B.~Melic, {\it {QCD light cone sum rule
  estimate of charming penguin contributions in $B \to \pi \pi$ }},  {\em Phys.
  Lett. B} {\bf 571} (2003) 75--84,
  [\href{http://arxiv.org/abs/hep-ph/0304179}{{\tt hep-ph/0304179}}].

\bibitem{Lubicz:2017syv}
{\bf ETM} Collaboration, V.~Lubicz, L.~Riggio, G.~Salerno, S.~Simula, and
  C.~Tarantino, {\it {Scalar and vector form factors of $D \to \pi(K) \ell \nu$
  decays with $N_f=2+1+1$ twisted fermions}},  {\em Phys. Rev. D} {\bf 96}
  (2017), no.~5 054514, [\href{http://arxiv.org/abs/1706.03017}{{\tt
  arXiv:1706.03017}}]. [Erratum: Phys.Rev.D 99, 099902 (2019), Erratum:
  Phys.Rev.D 100, 079901 (2019)].

\bibitem{Na:2010uf}
H.~Na, C.~T.~H. Davies, E.~Follana, G.~P. Lepage, and J.~Shigemitsu, {\it {The
  $D \rightarrow K, l \nu$ Semileptonic Decay Scalar Form Factor and $|V_{cs}|$
  from Lattice QCD}},  {\em Phys. Rev. D} {\bf 82} (2010) 114506,
  [\href{http://arxiv.org/abs/1008.4562}{{\tt arXiv:1008.4562}}].

\bibitem{Na:2011mc}
H.~Na, C.~T.~H. Davies, E.~Follana, J.~Koponen, G.~P. Lepage, and
  J.~Shigemitsu, {\it {$D \rightarrow \pi, l \nu$ Semileptonic Decays,
  $|V_{cd}|$ and 2$^{nd}$ Row Unitarity from Lattice QCD}},  {\em Phys. Rev. D}
  {\bf 84} (2011) 114505, [\href{http://arxiv.org/abs/1109.1501}{{\tt
  arXiv:1109.1501}}].

\bibitem{FLAG}
{\bf Flavour Lattice Averaging Group} Collaboration, S.~Aoki et~al., {\it {FLAG
  Review 2019: Flavour Lattice Averaging Group (FLAG)}},  {\em Eur. Phys. J. C}
  {\bf 80} (2020), no.~2 113, [\href{http://arxiv.org/abs/1902.08191}{{\tt
  arXiv:1902.08191}}].

\bibitem{Khodjamirian:2003xk}
A.~Khodjamirian, T.~Mannel, and M.~Melcher, {\it {Flavor SU(3) symmetry in
  charmless B decays}},  {\em Phys. Rev. D} {\bf 68} (2003) 114007,
  [\href{http://arxiv.org/abs/hep-ph/0308297}{{\tt hep-ph/0308297}}].

\bibitem{Duplancic:2008ix}
G.~Duplancic, A.~Khodjamirian, T.~Mannel, B.~Melic, and N.~Offen, {\it
  {Light-cone sum rules for $B \to \pi$ form factors revisited}},  {\em JHEP}
  {\bf 04} (2008) 014, [\href{http://arxiv.org/abs/0801.1796}{{\tt
  arXiv:0801.1796}}].

\bibitem{Piscopo:2023opf}
M.~L. Piscopo and A.~V. Rusov, {\it {Non-factorisable effects in the decays $
  {\overline{B}}_s^0\to {D}_s^{+}{\pi}^{-} $ and $ {\overline{B}}^0\to
  {D}^{+}{K}^{-} $ from LCSR}},  {\em JHEP} {\bf 10} (2023) 180,
  [\href{http://arxiv.org/abs/2307.07594}{{\tt arXiv:2307.07594}}].

\bibitem{Khodjamirian:2017fxg}
A.~Khodjamirian and A.~V. Rusov, {\it {$B_{s}\to K \ell \nu_\ell$ and $B_{(s)}
  \to \pi (K) \ell^+\ell^-$ decays at large recoil and CKM matrix elements}},
  {\em JHEP} {\bf 08} (2017) 112, [\href{http://arxiv.org/abs/1703.04765}{{\tt
  arXiv:1703.04765}}].

\bibitem{Cheng:2020vwr}
S.~Cheng, A.~Khodjamirian, and A.~V. Rusov, {\it {Pion light-cone distribution
  amplitude from the pion electromagnetic form factor}},  {\em Phys. Rev. D}
  {\bf 102} (2020), no.~7 074022, [\href{http://arxiv.org/abs/2007.05550}{{\tt
  arXiv:2007.05550}}].

\bibitem{Chetyrkin:2007vm}
K.~G. Chetyrkin, A.~Khodjamirian, and A.~A. Pivovarov, {\it {Towards NNLO
  Accuracy in the QCD Sum Rule for the Kaon Distribution Amplitude}},  {\em
  Phys. Lett. B} {\bf 661} (2008) 250--258,
  [\href{http://arxiv.org/abs/0712.2999}{{\tt arXiv:0712.2999}}].

\bibitem{Ball:2007zt}
P.~Ball, V.~M. Braun, and A.~Lenz, {\it {Twist-4 distribution amplitudes of the
  $K^*$ and $\phi$ mesons in QCD}},  {\em JHEP} {\bf 08} (2007) 090,
  [\href{http://arxiv.org/abs/0707.1201}{{\tt arXiv:0707.1201}}].

\bibitem{Khodjamirian:2006st}
A.~Khodjamirian, T.~Mannel, and N.~Offen, {\it {Form-factors from light-cone
  sum rules with B-meson distribution amplitudes}},  {\em Phys. Rev. D} {\bf
  75} (2007) 054013, [\href{http://arxiv.org/abs/hep-ph/0611193}{{\tt
  hep-ph/0611193}}].

\bibitem{Khodjamirian:2009ys}
A.~Khodjamirian, C.~Klein, T.~Mannel, and N.~Offen, {\it {Semileptonic charm
  decays $D \to \pi \ell \nu_\ell$ and $D \to K \ell \nu_\ell$ from QCD
  Light-Cone Sum Rules}},  {\em Phys. Rev. D} {\bf 80} (2009) 114005,
  [\href{http://arxiv.org/abs/0907.2842}{{\tt arXiv:0907.2842}}].

\bibitem{Charles:2004jd}
{\bf CKMfitter Group} Collaboration, J.~Charles, A.~Hocker, H.~Lacker,
  S.~Laplace, F.~R. Le~Diberder, J.~Malcles, J.~Ocariz, M.~Pivk, and L.~Roos,
  {\it {CP violation and the CKM matrix: Assessing the impact of the asymmetric
  $B$ factories}},  {\em Eur. Phys. J. C} {\bf 41} (2005), no.~1 1--131,
  [\href{http://arxiv.org/abs/hep-ph/0406184}{{\tt hep-ph/0406184}}].

\bibitem{Workman:2022ynf}
{\bf Particle Data Group} Collaboration, R.~L. Workman et~al., {\it {Review of
  Particle Physics}},  {\em PTEP} {\bf 2022} (2022) 083C01.

\bibitem{Leutwyler:1996qg}
H.~Leutwyler, {\it {The Ratios of the light quark masses}},  {\em Phys. Lett.
  B} {\bf 378} (1996) 313--318,
  [\href{http://arxiv.org/abs/hep-ph/9602366}{{\tt hep-ph/9602366}}].

\bibitem{Herren:2017osy}
F.~Herren and M.~Steinhauser, {\it {Version 3 of RunDec and CRunDec}},  {\em
  Comput. Phys. Commun.} {\bf 224} (2018) 333--345,
  [\href{http://arxiv.org/abs/1703.03751}{{\tt arXiv:1703.03751}}].

\bibitem{UTfit:2006vpt}
{\bf UTfit} Collaboration, M.~Bona et~al., {\it {The Unitarity Triangle Fit in
  the Standard Model and Hadronic Parameters from Lattice QCD: A Reappraisal
  after the Measurements of $\Delta m_s$ and ${\rm BR}(B \to \tau \nu_\tau)$}},
   {\em JHEP} {\bf 10} (2006) 081,
  [\href{http://arxiv.org/abs/hep-ph/0606167}{{\tt hep-ph/0606167}}].

\bibitem{Braun:1999uj}
V.~M. Braun, A.~Khodjamirian, and M.~Maul, {\it {Pion form-factor in QCD at
  intermediate momentum transfers}},  {\em Phys. Rev. D} {\bf 61} (2000)
  073004, [\href{http://arxiv.org/abs/hep-ph/9907495}{{\tt hep-ph/9907495}}].

\bibitem{Bijnens:2002mg}
J.~Bijnens and A.~Khodjamirian, {\it {Exploring light cone sum rules for pion
  and kaon form-factors}},  {\em Eur. Phys. J. C} {\bf 26} (2002) 67--79,
  [\href{http://arxiv.org/abs/hep-ph/0206252}{{\tt hep-ph/0206252}}].

\bibitem{Hiller:2012xm}
G.~Hiller, M.~Jung, and S.~Schacht, {\it {SU(3)-flavor anatomy of nonleptonic
  charm decays}},  {\em Phys. Rev. D} {\bf 87} (2013), no.~1 014024,
  [\href{http://arxiv.org/abs/1211.3734}{{\tt arXiv:1211.3734}}].

\bibitem{Hiller:2013awa}
G.~Hiller, M.~Jung, and S.~Schacht, {\it {SU(3)$_{F}$ in nonleptonic charm
  decays}},  {\em PoS} {\bf EPS-HEP2013} (2013) 371,
  [\href{http://arxiv.org/abs/1311.3883}{{\tt arXiv:1311.3883}}].

\bibitem{Muller:2015lua}
S.~M\"uller, U.~Nierste, and S.~Schacht, {\it {Topological amplitudes in $D$
  decays to two pseudoscalars: A global analysis with linear $SU(3)_F$
  breaking}},  {\em Phys. Rev. D} {\bf 92} (2015), no.~1 014004,
  [\href{http://arxiv.org/abs/1503.06759}{{\tt arXiv:1503.06759}}].

\bibitem{Gavrilova:2023fzy}
M.~Gavrilova, Y.~Grossman, and S.~Schacht, {\it {Determination of the
  $D\rightarrow \pi\pi$ Penguin over Tree Ratio}},
  \href{http://arxiv.org/abs/2312.10140}{{\tt arXiv:2312.10140}}.

\end{thebibliography}\endgroup

\end{document}